\documentclass[11pt]{article}
\usepackage{amssymb}

\evensidemargin\oddsidemargin \topmargin2cm

\newtheorem{theorem}{Theorem}
\newtheorem{lemma}{Lemma}
\newtheorem{corollary}[lemma]{Corollary}
\newtheorem{remark}{Remark}
\newenvironment{proof}[1][Proof]{\textbf{#1.} }{\ \rule{0.5em}{0.5em}}

\begin{document}

\bigskip

\begin{center}
{\LARGE Superselection rules induced by infrared divergence}

\bigskip

{\large Joachim Kupsch}\footnote{%
e-mail: kupsch@physik.uni-kl.de}

{\large Fachbereich Physik, TU Kaiserslautern\\[0pt]
D-67653 Kaiserslautern, Germany}

\bigskip
\end{center}

\begin{abstract}
Superselection rules induced by the interaction with a mass zero
Boson field are investigated for a class of exactly soluble
Hamiltonian models. The calculations apply as well to discrete as
to continuous superselection rules. The initial state (reference
state) of the Boson field is either a normal state or a KMS state.
The superselection sectors emerge if and only if the Boson field
is infrared divergent, i. e. the bare photon number diverges and
the ground state of the Boson field disappears in the continuum.
The time scale of the decoherence depends on the strength of the
infrared contributions of the interaction and on properties of the
initial state of the Boson system. These results are first derived
for a Hamiltonian with conservation laws. But in the most general
case the Hamiltonian includes an additional scattering potential,
and the only conserved quantity is the energy of the total system.
The superselection sectors remain stable against the perturbation
by the scattering processes.
\end{abstract}

\section{Introduction}

Superselection rules are the basis for the emergence of classical physics
within quantum theory. But despite of the great progress in understanding
superselection rules, see e.g. \cite{Wightman:1995}, quantum mechanics and
quantum field theory do not provide enough exact superselection rules to
infer the classical probability of ``facts'' from quantum theory. This
problem is most often discussed in the context of measurement of quantum
mechanical objects. In an important paper about the process of measurement
Hepp \cite{Hepp:1972} has presented a class of models for which the dynamics
induces superselection sectors. Hepp starts with a very large algebra of
observables -- essentially all observables with the exception of the
``observables at infinity'' which constitute an a priory set of
superselection rules -- and the superselection sectors emerge in the weak
operator convergence. But it has soon been realized that the algebra of
observables, which is relevant for the understanding of the process of
measurement \cite{Emch:1972a} \cite{Araki:1980} and, more generally for the
understanding of the classical appearance of the world \cite{Zurek:1982}
\cite{Joos/Zeh:1985} \cite{JZKGKS:2003} can be severely restricted. Then
strong or even uniform operator convergence is possible.

A system, which is weakly coupled to an environment, which has a Hamiltonian
with a continuous spectrum, usually decays into its ground state, if the
environment is in a normal state; or the system approaches a canonical
ensemble, if the environment is in a state with positive temperature. More
interesting decoherence effects may occur on an intermediate time scale, or
in systems, for which the decay or the thermalization are prevented by
conservation laws. To emphasize effects on an intermediate time scale one
can use a strong coupling between system and environment. This method has
some similarity to the singular coupling method of the Markov approximation,
which also scales the dynamics at an intermediate time period to large
times. The basic model, which we discuss, has therefore the following
properties: existence of conservation laws and strong coupling. Thereby
strong coupling means that the spectral properties of the Hamiltonian are
modified by the interaction term.

In this paper, which is an extension of \cite{Kupsch:2000a}, we investigate
the emergence of superselection rules for a system, which is coupled to a
mass zero Boson field. The dynamics of the total system is always generated
by a semibounded Hamiltonian. The restriction to the Boson sector
corresponds to a van Hove model \cite{Hove:1952}. As the main result we
prove for a class of such models:\newline
-- The superselection rules are induced by the infrared contributions of the
Boson field.\newline
-- The superselection sectors are stable for $t\rightarrow \infty $ if and
only if the Boson field is infrared divergent.

The infrared divergence of the van Hove model has been studied by Schroer
\cite{Schroer:1963} more than forty years ago. The Boson field is still
defined on the Fock space, but the ground state of the Boson field
disappears in the continuum. In the usual discussions of decoherence this
type of infrared divergence corresponds to the ohmic or subohmic case \cite
{LCD:1987}. As additional result we prove that the induced superselection
sectors are stable against perturbation by scattering processes.

The paper is organized as follows. In Sect. \ref{ssr} we give a short
introduction to the dynamics of subsystems and to superselection rules
induced by the environment. The calculations are preformed in the
Schr\"{o}dinger picture, which allows also non-factorized initial states. We
prove that the off-diagonal matrix elements of the reduced statistical
operator can be suppressed in trace norm for discrete and for continuous
superselection rules. In Sect. \ref{field} we investigate a class of
Hamiltonian models with the environment given by a mass zero Boson field,
and the interplay between infrared divergence and induced superselection
rules is derived. The resulting superselection sectors do not depend on the
initial state; they finally emerge for all initial states of the total
system. But to have superselection sectors, which are effective on a short
time scale, the reference state of the environment has to satisfy some
``smoothness'' conditions.

In Sect. \ref{KMS} we admit a KMS state of positive temperature as reference
state of the Boson system. Again the same superselection sectors emerge,
even on a shorter time scale. In the final Sect. \ref{scattering} we prove
that the induced superselection sectors are stable against additional
scattering processes. Some technical details for the Sects. \ref{ssr} and
\ref{field} are given in the Appendices \ref{estimates} and \ref{vanHove}.

\section{Induced superselection rules\label{ssr}}

\subsection{General considerations}

We start with a few mathematical notations. Let $\mathcal{H}$ be a separable
Hilbert space, then the following spaces of linear operators are used.
\newline
$\mathcal{B}(\mathcal{H})$: The linear space of all bounded operators $A$
with the operator norm $\Vert A\Vert $.\newline
$\mathcal{T}(\mathcal{H})$: The linear space of all nuclear operators $A$
with the trace norm $\Vert A\Vert _{1}=\mathrm{tr}\sqrt{A^{+}A}$.\newline
$\mathcal{S}(\mathcal{H})$: The set of all positive nuclear operators $W$
with a normalized trace, $\mathrm{tr}\,W=1$.\newline
If $A$ is a closed (unbounded) linear operator, then $\mathcal{D}(A)\subset
\mathcal{H}$ denotes the domain of definition of this operator.

With the exception of Sect. \ref{KMS}, where also KMS states are admitted
for the environment, we assume standard quantum mechanics where any state of
a quantum system is represented by a statistical operator $W\in \mathcal{S}(%
\mathcal{H})$; the rank one projection operators thereby correspond to the
pure states. Without additional knowledge about the structure of the system
we have to assume that the set of all states corresponds to $\mathcal{S}(%
\mathcal{H})$, and the operator algebra of all (bounded) observables
coincides with $\mathcal{B}(\mathcal{H})$.

In the following we consider an \textit{open system}, i.e. a system $S$
which interacts with an environment $E$, such that the total system $S\times
E$ satisfies the usual Hamiltonian dynamics. The Hilbert space $\mathcal{H}%
_{S\times E}$ of the total system is the tensor space $\mathcal{H}%
_{S}\otimes \mathcal{H}_{E}$ of the Hilbert spaces for $S$ and for $E$. Let $%
W\in \mathcal{S}(\mathcal{H}_{S\times E})$ be the state of the total system
and $A\in \mathcal{B}(\mathcal{H}_{S})$ be an observable of the system $S$,
then the expectation $\mathrm{tr}_{S\times E}W(A\otimes I_{E})$ satisfies
the identity $\mathrm{tr}_{S\times E}\,(A\otimes I_{E})W=\mathrm{tr}%
_{S}\,A\rho $ with the reduced statistical operator $\rho _{S}=\mathrm{tr}%
_{E}W\in \mathcal{S}(\mathcal{H}_{S})$. Here the symbols $\mathrm{tr}_{S}$, $%
\mathrm{tr}_{E}$ and $\mathrm{tr}_{S\times E}$ denote the (partial) traces
with respect to the Hilbert spaces $\mathcal{H}_{S}$, $\mathcal{H}_{E}$ or $%
\mathcal{H}_{S\times E}$, respectively. We shall refer to $\rho _{S}=\mathrm{%
tr}_{E}W$ as the \textit{state} of the system $S$. As indicated above we
consider the usual Hamiltonian dynamics for the total system, i.e. $%
W\rightarrow W(t)=U(t)WU^{+}(t)\in \mathcal{S}(\mathcal{H}_{S\times E})$
with the unitary group $U(t)=\exp (-iH_{S\times E}t)$ generated by the total
Hamiltonian $H_{S\times E}$. Except for the trivial case that $S$ and $E$ do
not interact, the dynamics of the reduced statistical operator $\rho _{S}(t)=%
\mathrm{tr}_{E}\,U(t)WU^{+}(t)$ does no longer follow a group law; and it is
exactly this dynamics which can produce induced superselection sectors.

In order to define a linear dynamics $\rho _{S}\rightarrow \rho _{S}(t)$ for
the state of the system $S$ we have to assume that the initial state
factorizes as
\begin{equation}
W=\rho _{S}\otimes \rho _{E},  \label{2.1}
\end{equation}
see \cite{KSS:2001}. Here $\rho _{S}\in \mathcal{S}(\mathcal{H}_{S})$ is the
initial state of the system and $\rho _{E}\in \mathcal{S}(\mathcal{H}_{E})$
is the reference state of the environment. The dynamics of the reduced
statistical operator $\rho _{S}(t)$ then follows as
\begin{equation}
\rho _{S}\in \mathcal{S}(\mathcal{H}_{S})\rightarrow \rho _{S}(t)=\Phi
_{t}(\rho _{S}):=\mathrm{tr}_{E}U(t)\left( \rho _{S}\otimes \rho _{E}\right)
U^{\dagger }(t)\in \mathcal{S}(\mathcal{H}_{S}).  \label{2.2}
\end{equation}
The reduced dynamics $\Phi _{t}(\rho _{S})$ can be extended to a continuous
linear mapping $\rho _{S}\in \mathcal{T}(\mathcal{H}_{S})\rightarrow \Phi
_{t}(\rho _{S})\in \mathcal{T}(\mathcal{H}_{S})$ with the obvious properties
\begin{equation}
\left\| \Phi _{t}(\rho _{S})\right\| _{1}\leq \left\| \rho _{S}\right\|
_{1},~\mathrm{tr}_{S}\Phi _{t}(\rho _{S})=\mathrm{tr}_{S}\rho _{S},~\Phi
_{t}(\rho _{S})\geq 0\;\mathrm{if}\;\rho _{S}\geq 0.  \label{2.3}
\end{equation}
Here $\left\| \cdot \right\| _{1}$ is the trace norm of operators on $%
\mathcal{H}_{S}$.

Discrete and continuous superselection rules are characterized by
a self-adjoint superselection operator $F=\int_{\mathbb{R}}\lambda
P(d\lambda )$. The projection operators $P(\Delta )$ of the
spectral resolution of this operator are defined for all intervals
$\Delta =\left[ a,b\right) $ of the real line and satisfy
\begin{equation}
\begin{array}{c}
P(\Delta ^{1}\cup \Delta ^{2})=P(\Delta ^{1})+P(\Delta ^{2})\mathrm{\quad
if\quad }\Delta ^{1}\cap \Delta ^{2}=\emptyset \\
P(\Delta ^{1})P(\Delta ^{2})=P(\Delta ^{1}\cap \Delta
^{2}),\,P(\emptyset )=0,\,P(\mathbb{R})=1.
\end{array}
\label{2.4}
\end{equation}
The mapping $\Delta \rightarrow P(\Delta )$ can be extended to the $\sigma $%
-algebra of the real line $\mathcal{B}(\mathbb{R})$ generated by
open sets. In the most commonly discussed case of discrete
superselection rules the function $x\in \mathbb{R}\rightarrow
P(\left( -\infty ,x\right) )$ is a step function. In the case of
continuous superselection rules, in which we are mainly interested
in, the function $x\in \mathbb{R}\rightarrow P(\left( -\infty
,x\right) )$ is strongly continuous, and the projection operators
$P\left( \left( a,b\right) \right) =P\left( \left[ a,b\right)
\right) =P\left( \left[ a,b\right] \right) $ coincide.

The dynamics of the total system $S\times E$ induces superselection rules
into the system $S$, if there exists a family of projection operators $%
\left\{ P_{S}(\Delta )\mid \Delta \subset \mathbb{R}\right\} $ on
the Hilbert space $\mathcal{H}_{S}$, which satisfies the rules
(\ref{2.4}), such that the off-diagonal parts $P_{S}(\Delta
^{1})\Phi _{t}(\rho _{S})P_{S}(\Delta ^{2})$ of the statistical
operators of the system $S$ are dynamically suppressed, i.e.
\newline $P_{S}(\Delta ^{1})\Phi _{t}(\rho _{S})P_{S}(\Delta
^{2})\rightarrow 0$\ if\ $\,t\rightarrow \infty $\ and\
$\mathrm{dist}(\Delta ^{1},\Delta ^{2})>0$. In the subsequent
sections we derive superselection rules, for which the
off-diagonal parts of the statistical operator even vanish in
trace norm
\begin{equation}
\left\| P_{S}(\Delta ^{1})\Phi _{t}(\rho _{S})P_{S}(\Delta ^{2})\right\|
_{1}\rightarrow 0\quad \mathrm{if}\;t\rightarrow \infty  \label{2.5}
\end{equation}
for all initial states $\rho _{S}\in \mathcal{S}(\mathcal{H}_{S})$ and all
separated intervals $\Delta ^{1}$ and $\Delta ^{2}$. This statement, which
does not specify the time scale of the decoherence process, can be used as
definition of induced superselection rules. But to have superselection
rules, which contribute to the emergence of classical properties, the
decrease of (\ref{2.5}) has to be sufficiently fast. We shall come back to
that problem later.

\subsection{Models}

For all models we are investigating, the total Hamiltonian is defined on the
tensor space $\mathcal{H}_{S\times E}=\mathcal{H}_{S}\otimes \mathcal{H}_{E}$
as
\begin{eqnarray}
H_{S\times E} &=&H_{S}\otimes I_{E}+I_{S}\otimes H_{E}+F\otimes G  \nonumber
\\
&=&\left( H_{S}-\frac{1}{2}F^{2}\right) \otimes I_{E}+\frac{1}{2}\left(
F\otimes I_{E}+I_{S}\otimes G\right) ^{2}+I_{S}\otimes \left( H_{E}-\frac{1}{%
2}G^{2}\right)  \label{mod.3}
\end{eqnarray}
where $H_{S}$ is the positive Hamiltonian of $S$, $H_{E}$ is the positive
Hamiltonian of $E$, and $F\otimes G$ is the interaction potential between $S$
and $E$ with operators $F$ on $\mathcal{H}_{S}$ and $G$ on $\mathcal{H}_{E}$%
. To guarantee that $H_{S\times E}$ is self-adjoint and semibounded we assume

\begin{enumerate}
\item[1)]  The operators $F$ and $F^{2}$ ($G$ and $G^{2})$ are essentially
self-adjoint on the domain of $H_{S}$ ($H_{E}$). The operators $H_{S}-\frac{1%
}{2}F^{2}$ and $H_{E}-\frac{1}{2}G^{2}$ are semibounded.
\end{enumerate}

Since $F^{2}\otimes I_{E}\pm 2F\otimes G+I_{S}\otimes G^{2}$ are positive
operators, the operator $F\otimes G$ is \newline
$\left( H_{S}\otimes I_{E}+I_{S}\otimes H_{E}\right) $-bounded with relative
bound one, and W\"{u}st's theorem, see e.g. Theorem X.14 in \cite
{Reed/Simon:1975}, implies that $H_{S\times E}$ is essentially self-adjoint
on the domain of $H_{S}\otimes I_{E}+I_{S}\otimes H_{E}$. Moreover $%
H_{S\times E}$ is obviously semibounded.

To derive induced superselection rules we need the rather severe restriction

\begin{enumerate}
\item[2)]  The operators $H_{S}$ and $F$ commute strongly, i.e. their
spectral projections commute.
\end{enumerate}

This assumption implies that $F$ is a conserved quantity of the dynamics
generated by the Hamiltonian (\ref{mod.3}). The operator $F$ has a spectral
representation
\begin{equation}
F=\int_{\mathbb{R}}\lambda P_{S}(d\lambda )  \label{mod.4}
\end{equation}
with a family (\ref{2.4}) of projection operators $P_{S}(\Delta )$
indexed by measurable subsets $\Delta \subset \mathbb{R}$. We
shall see below that exactly the projection operators of this
spectral representation determine the induced superselection
sectors.

As a consequence of assumption 2) we have $\left[ H_{S},P_{S}(\Delta )\right]
=0$ for all intervals $\Delta \subset \mathbb{R}$. The Hamiltonian (\ref{mod.3}%
) has therefore the form $H_{S\times E}=H_{S}\otimes I_{E}+\int_{\mathbb{R}%
}P_{S}(d\lambda )\otimes \left( H_{E}+\lambda G\right) $. The operator $%
\left| G\right| =\sqrt{G^{2}}$ has the upper bound $\left| G\right| \leq
aG^{2}+(4a)^{-1}I$ with an arbitrarily small constant $a>0$. Since $G^{2}$
is $H_{E}$-bounded with relative bound $2$, the operator $G$ is $H_{E}$%
-bounded with an arbitrarily small bound. The Kato-Rellich
theorem, see e.g. \cite{Reed/Simon:1975}, implies that the
operators $H_{E}+\lambda G$ are self-adjoint on the domain of
$H_{E}$ for all $\lambda \in \mathbb{R}$. The unitary evolution
$U(t):=\exp (-iH_{S\times E}t)$ of the total system can therefore
be written as \newline $U(t)=\left( U_{S}(t)\otimes I_{E}\right)
\int dP_{S}(\lambda )\otimes \exp \left( -i\left( H_{E}+\lambda
G\right) t\right) $, where
\begin{equation}
U_{S}(t)=\exp \left( -iH_{S}t\right)  \label{mod.5}
\end{equation}
is the unitary evolution of the system $S$. The evolution (\ref{2.2}) of an
initial state $\rho _{S}\in \mathcal{S}(\mathcal{H}_{S})$ follows as
\begin{equation}
\Phi _{t}(\rho _{S})=U_{S}(t)\left( \int_{\mathbb{R}\times
\mathbb{R}}\chi \left( \alpha ,\beta ;t\right) P_{S}(d\alpha
)\,\rho _{S}\,P_{S}(d\beta )\right) U_{S}^{+}(t)  \label{mod.10}
\end{equation}
with the trace
\begin{equation}
\chi (\alpha ,\beta ;t)=\mathrm{tr}_{E}\left( \mathrm{e}^{i\left(
H_{E}+\alpha G\right) t}\mathrm{e}^{-i\left( H_{E}+\beta G\right) t}\rho
_{E}\right) .  \label{mod.11}
\end{equation}
For the models investigated in Sect. \ref{field} this trace factorizes into
\begin{equation}
\chi (\alpha ,\beta ;t)=\mathrm{e}^{i\vartheta (\alpha ,t)}\chi _{0}(\alpha
-\beta ;t)\mathrm{e}^{-i\vartheta (\beta ,t)}  \label{mod.12}
\end{equation}
where $\vartheta (\alpha ,t)$ is a real phase. The function $\chi
_{0}(\lambda ;t)=\overline{\chi _{0}(-\lambda ;t)}$ and its derivative can
be estimated by
\begin{equation}
\left| \chi _{0}(\lambda ;t)\right| \leq \phi \left( \lambda ^{2}\zeta
(t)\right) ,\quad \int_{\delta }^{\infty }\left| \frac{\partial }{\partial
\lambda }\chi _{0}(\lambda ;t)\right| d\lambda \leq \phi \left( \delta
^{2}\zeta (t)\right) \;\mathrm{for~all}\;\delta \geq 0.  \label{mod.13}
\end{equation}
Thereby $\phi (s)$ is a positive non increasing function which vanishes for $%
s\rightarrow \infty $ such that $\int^{\infty }\phi (s)ds<\infty $, and the
time dependent positive function $\zeta (t)$ increases to infinity$\,$ if $%
t\rightarrow \infty $. The factorization (\ref{mod.11}) implies
that the operator (\ref{mod.10}) is the product\newline $\Phi
_{t}(\rho _{S})=U_{S}(t)U_{\vartheta }(t)\left(
\int_{\mathbb{R}\times \mathbb{R}}\chi _{1}\left( \alpha -\beta
;t\right) P_{S}(d\alpha )\,\rho _{S}\,P_{S}(d\beta )\right)
U_{\vartheta }^{+}(t)U_{S}^{+}(t)$ with the unitary operator
$U_{\vartheta }(t)=\int \exp \left( i\vartheta (\alpha ,t)\right)
P_{S}(d\alpha )$. As the projection operators $P_{S}(\Delta )$
commute with the unitary operators $U_{S}(t)$ and $U_{\vartheta
}(t)$, the trace norm of $P_{S}(\Delta ^{1})\Phi _{t}(\rho
_{S})P_{S}(\Delta ^{2})$ is given by
\begin{equation}
\left\| P_{S}(\Delta ^{1})\Phi _{t}(\rho _{S})P_{S}(\Delta ^{2})\right\|
_{1}=\left\| \int_{\Delta ^{1}\times \Delta ^{2}}\chi _{0}\left( \alpha
-\beta ;t\right) P_{S}(d\alpha )\,\rho _{S}\,P_{S}(d\beta )\right\| _{1}.
\label{mod.14}
\end{equation}
The phase function does not contribute to this norm. In Appendix \ref
{estimates} we prove that the estimate (\ref{mod.13}) is sufficient to
derive the upper bound
\begin{equation}
\left\| P_{S}(\Delta ^{1})\Phi _{t}(\rho _{S})P_{S}(\Delta ^{2})\right\|
_{1}\leq \phi \left( \delta ^{2}\zeta (t)\right)  \label{mod.16}
\end{equation}
for arbitrary intervals $\Delta ^{1}$ and $\Delta ^{2}$ with a distance $%
\delta \geq 0$. This estimate is uniform for all initial states $\rho
_{S}\in \mathcal{S}(\mathcal{H}_{S})$. The arguments of Appendix \ref
{estimates} are applicable to superselection operators (\ref{mod.4}) $F$
with an arbitrary spectrum. For operators $F=\sum \lambda _{n}P_{n}^{S}$
with a discrete spectrum, which has no accumulation point, uniform norm
estimates can be derived with simpler methods, see \cite{Kupsch:2000} or
Sect. 7.6 of \cite{JZKGKS:2003}.

The norm (\ref{mod.16}) vanishes for all intervals with a positive distance $%
\mathrm{dist}(\Delta ^{1},\Delta ^{2})=\delta >0$ on a time scale, which
depends on the functions $\zeta (t)$ and $\phi (s)$. The function $\zeta (t)$
is mainly determined by the Hamiltonian, whereas $\phi (s)$ depends strongly
on the reference state of the environment.

\begin{remark}
\label{AZ}A simple class of explicitly soluble models, which yield estimates
similar to (\ref{mod.16}), can be obtained under the additional assumptions%
\newline
-- the operator $G$ has an absolutely continuous spectrum,\newline
-- the Hamiltonian $H_{E}$ and the potential $G$ commute strongly.\newline
Models of this type have been investigated (for operators $F$ with a
discrete spectrum) by Araki \cite{Araki:1980} and by Zurek \cite{Zurek:1982}%
, see also Sect. 7.6 of \cite{JZKGKS:2003} and \cite{Kupsch:2000}.
With these additional assumptions the trace (\ref{mod.11})
simplifies to $\chi (\alpha ,\beta ;t)=\mathrm{tr}_{E}\left(
\mathrm{e}^{i(\alpha -\beta )Gt}\rho _{E}\right) $. Let
$G=\int_{\mathbb{R}}\lambda P_{E}(d\lambda )$ be the spectral
representation of the operator $G$. Then the measure $d\mu
(\lambda ):=\mathrm{tr}_{E}\left( P_{E}(d\lambda )\,\rho
_{E}\right) $ is absolutely continuous with respect to the
Lebesgue measure for any $\rho
_{E}\in \mathcal{S}(\mathcal{H}_{E})$, and the function $\chi (t):=\mathrm{tr%
}_{E}\left( \mathrm{e}^{iGt}\rho _{E}\right) =\int_{\mathbb{R}}\mathrm{e}%
^{i\lambda t}$ $d\mu (\lambda )$ vanishes for $t\rightarrow \infty $. Under
suitable restrictions on the reference state the measure $d\mu (\lambda )=%
\mathrm{tr}_{E}\left( P_{E}(d\lambda )\,\rho _{E}\right) $ has a smooth
density, and we can derive a fast decrease of the Fourier transform $\chi
(t) $ and its derivatives. That implies upper bounds similar to (\ref{mod.13}%
) and a fast decrease of (\ref{mod.11}) $\chi (\alpha ,\beta ;t)$ for $%
\alpha \neq \beta $.
\end{remark}

\begin{remark}
\label{Heisenberg}Instead of the dynamics (\ref{2.2}) in the Schr\"{o}dinger
picture we can use the Heisenberg dynamics
\begin{equation}
A\in \mathcal{B}(\mathcal{H}_{S})\rightarrow \Psi _{t}(A)=\Psi _{t}(A):=%
\mathrm{tr}_{E}U^{\dagger }(t)\left( A\otimes I_{E}\right) U(t)\rho _{E}\in
\mathcal{B}(\mathcal{H}_{S})  \label{mod.17}
\end{equation}
to investigate induced superselection rules. As the estimate (\ref{mod.16})
is uniform with respect to the initial state $\rho _{S}\in \mathcal{S}(%
\mathcal{H}_{S})$, the duality relation $\mathrm{tr}_{S}\left( P_{S}(\Delta
^{1})\Phi _{t}(\rho _{S})P_{S}(\Delta ^{2})A\right) =\newline
\mathrm{tr}_{S}\,\rho _{S}\Psi _{t}\left( P_{S}(\Delta ^{2})AP_{S}(\Delta
^{1})\right) $ leads to a criterion for induced superselection rules in the
Heisenberg picture:
\begin{equation}
\lim_{t\rightarrow \infty }\left\| \Psi _{t}\left( P_{S}(\Delta
^{1})AP_{S}(\Delta ^{2})\right) \right\| =0  \label{mod.18}
\end{equation}
for all observables $A\in \mathcal{B}(\mathcal{H}_{S})$ and for all
intervals $\Delta ^{1}$ and $\Delta ^{2}$ with a distance \newline
$\mathrm{dist}(\Delta ^{1},\Delta ^{2})>0$. In the case of models with the
Hamiltonian (\ref{mod.3}) which satisfy Assumption 2), the condition (\ref
{mod.18}) is equivalent to a more transparent condition. For these models
the full dynamics $U(t)=\exp (-iH_{S\times E}t)$ commutes with $P_{S}(\Delta
)\otimes I_{E}$, and the Heisenberg dynamics (\ref{mod.17}) satisfies the
identities $P_{S}(\Delta )\Psi _{t}(A)=\Psi _{t}(P_{S}(\Delta )A)\;$and$%
\;\Psi _{t}(A)P_{S}(\Delta )=\Psi _{t}(AP_{S}(\Delta ))$. These identities
and (\ref{mod.18}) imply that the off-diagonal parts of $\Psi _{t}(A)$ have
to vanish for all observables $A\in \mathcal{B}(\mathcal{H}_{S})$ and for
all disjoint intervals with a non-vanishing distance at large $t$%
\begin{equation}
\lim_{t\rightarrow \infty }\left\| P_{S}(\Delta ^{1})\Psi _{t}\left(
A\right) P_{S}(\Delta ^{2})\right\| =0.  \label{mod.19}
\end{equation}
This criterion resembles the definition of the exact superselection rules:
\newline
$P_{S}(\Delta ^{1})\,A\,P_{S}(\Delta ^{2})=0$ for all $\Delta ^{1}\cap
\Delta ^{2}=\emptyset $, see e.g. \cite{Jauch:1960} or \cite{Wightman:1995}.
The criterion (\ref{mod.19}) has been used in \cite{Kupsch:2000a} to derive
induced superselection rules for the model of Sect. \ref{field}.
\end{remark}

\section{The interaction with a Boson field\label{field}}

\subsection{The Hamiltonian}

We choose a system $S$ which satisfies the constraints 1) and 2),
and the environment $E$ is given by a Boson field. As specific
example we may consider a spin system with Hilbert space
$\mathcal{H}_{S}=\mathbb{C}^{2}$ and
Hamiltonian $H_{S}=\alpha \sigma _{3}$ and $F=\beta \sigma _{3}$ where $%
\alpha \geq 0$ and $\beta $ are real constants and $\sigma _{3}$ is the
Pauli spin matrix. A more interesting example is a particle on the real line
with velocity coupling. The Hilbert space of the particle is $\mathcal{H}%
_{S}=\mathcal{L}^{2}(\mathbb{R})$. The Hamiltonian and the
interaction potential of the particle are
\begin{equation}
H_{S}=\frac{1}{2}P^{2}\;\mathrm{and}\;F=P  \label{field.1}
\end{equation}
where $P=-i\,d/dx$ is the momentum operator of the particle. The identity $%
H_{S}-\frac{1}{2}F^{2}=0$ guarantees the positivity of the first term in (%
\ref{mod.3}).

As Hilbert space $\mathcal{H}_{E}$ we choose the Fock space of symmetric
tensors $\mathcal{F}(\mathcal{H}_{1})$ based on the one particle Hilbert
space $\mathcal{H}_{1}$. The Hamiltonian is generated by a one-particle
Hamilton operator $M$ on$\,\mathcal{H}_{1}$ with the following properties

(i) $M\;$is\thinspace a\thinspace positive operator with an absolutely
continuous spectrum,

(ii) $M\;$has\thinspace an\thinspace unbounded\thinspace inverse$\,M^{-1}$.

The spectrum of $M$ is (a subset of) $\mathbb{R}_{+}$, which -- as
a consequence of the second assumption -- includes zero. The
Hamiltonian of the free field is then the derivation
$H_{E}=d\Gamma (M)$ generated by $M$,
see Appendix \ref{vanHove}. As explicit example we may take $\mathcal{H}_{1}=%
\mathcal{L}^{2}(\mathbb{R}^{n})$ with inner product $\left( f\mid
g\right) =\int_{\mathbb{R}^{n}}\overline{f(k)}g(k)d^{n}k$. The
one-particle Hamilton operator can be chosen as $\left( Mf\right)
(k):=\varepsilon (k)f(k)$ with the positive energy function
$\varepsilon (k)=c\left| k\right| ,\,c>0,\,k\in \mathbb{R}^{n}.$
Let $a_{k}^{\#},\,k\in \mathbb{R}^{n}$, denote the distributional
creation/annihilation operators, such that $a^{+}(f)=\int
a_{k}^{+}\,f(k)d^{n}k$ and $a(f)=\int
a_{k}\,\overline{f(k)}d^{n}k$ are the creation/annihilation
operators of the vector $f\in \mathcal{H}_{1}$, normalized to
$\left[ a(f),a^{+}(g)\right] =\left( f\mid g\right) $. The
Hamiltonian $H_{E}=d\Gamma (M)$ coincides with $H_{E}=\int
\varepsilon (k)a_{k}^{+}a_{k}d^{n}k$. The interaction potential
$G$ is chosen as the self-adjoint field operator
\begin{equation}
G=\Phi (h):=a^{+}(h)+a(h)  \label{field.2}
\end{equation}
where the vector $h\in \mathcal{H}_{1}$ satisfies the additional constraint
\begin{equation}
2\left\| M^{-\frac{1}{2}}h\right\| \leq 1.  \label{field.4}
\end{equation}
This constraint guarantees that $H_{E}-\frac{1}{2}\Phi ^{2}(h)$ is bounded
from below, and the Hamiltonian (\ref{mod.3}) is a well defined semibounded
operator on $\mathcal{H}_{S}\otimes \mathcal{F}(\mathcal{H}_{1})$, see
Appendix \ref{vanHove}. In the sequel we always assume that (\ref{field.4})
is satisfied.

To derive induced superselection sectors we have to estimate the time
dependence of the traces (\ref{mod.11}) $\chi (\alpha ,\beta ;t)=\mathrm{tr}%
_{E}U_{\alpha \beta }(t)\rho _{E}$ where $\rho _{E}$ is the reference state
of the Boson field, and the unitary operators $U_{\alpha \beta }(t)$ are
given by
\begin{equation}
U_{\alpha \beta }(t):=\exp (iH_{\alpha }t)\exp (-iH_{\beta }t),\;\mathrm{with%
}\;H_{\alpha }=H_{E}+\alpha \Phi (h),\;\alpha ,\beta \in
\mathbb{R}\mathbf{.} \label{field.5}
\end{equation}
The Hamiltonians $H_{\alpha }$ are Hamiltonians of the van Hove model \cite
{Hove:1952}. Details for the following statements are given in the Appendix
\ref{vanHove}. The Hamiltonian $H_{E}+\Phi (h)$ is defined on the Fock space
$\mathcal{F}(\mathcal{H}_{1})$ as semibounded self-adjoint operator if $h\in
\mathcal{H}_{1}$ is in the domain of $M^{-\frac{1}{2}}$, $h\in \mathcal{D}%
(M^{-\frac{1}{2}})$. But this Hamiltonian has a ground state only if the low
energy contributions of $h$ are not too strong, more precisely, if
\begin{equation}
h\in \mathcal{D}(M^{-1})  \label{field.6}
\end{equation}
is satisfied. Under this more restrictive condition the Hamiltonian has
another important property: $H_{E}+\Phi (h)$ is unitarily equivalent to the
free Hamiltonian $H_{E}$
\begin{equation}
H_{E}+\Phi (h)=T^{+}(M^{-1}h)H_{E}T(M^{-1}h)-\left\| M^{-\frac{1}{2}%
}h\right\| ^{2}.  \label{field.7}
\end{equation}
Thereby the intertwining operators are the unitary Weyl operators $T(f)=%
\newline
\exp \left( a^{+}(f)-a(f)\right) $ defined for $f\in \mathcal{H}_{1}$.

\subsection{Coherent states as reference state\label{coherent}}

For the further calculations we first choose as reference state a coherent
state. Let $f\in \mathcal{H}_{1}\rightarrow \exp f=1_{vac}+f+\frac{1}{2}%
f\circ f+..\in \mathcal{F}(\mathcal{H}_{1})$ be the convergent exponential
series of the symmetric tensor algebra of the Fock space. Thereby $%
1_{vac}\in \mathcal{F}(\mathcal{H}_{1})$ is the vacuum vector. Then $%
T(f)1_{vac}=\exp \left( f-\frac{1}{2}\left\| f\right\| ^{2}\right) $ is a
normalized exponential vector or coherent state. The reference state $\rho
_{E}$ is the projection operator $\omega (f)$ onto this vector, i.e. $\omega
(f)=T(f)P_{vac}T^{+}(f)$ where $P_{vac}$ is the projection operator onto the
vacuum. The basic identity which characterizes the coherent states is the
expectation of the Weyl operators
\begin{equation}
\mathrm{tr}_{E}T(h)\omega (f)=\exp \left( -\frac{1}{2}\left\| h\right\|
^{2}\right) \exp \left( 2i\,\mathrm{Im}\left( f\mid h\right) \right)
\label{field.20}
\end{equation}
Under the assumption (\ref{field.6}) the trace (\ref{mod.11}) is calculated
in Appendix \ref{vanHove} using (\ref{field.7}) and properties of the Weyl
operators. The result is
\begin{equation}
\mathrm{tr}_{E}U_{\alpha \beta }(t)\omega (f)=\exp \left( -\left( \alpha
-\beta \right) ^{2}\zeta (t)\right) \exp \left( i\left( \vartheta (\alpha
,t)-\vartheta (\beta ,t)\right) \right)  \label{field.8}
\end{equation}
with
\begin{equation}
\zeta (t)=\frac{1}{2}\left\| \left( I-\exp (iMt)\right) M^{-1}h\right\| ^{2};
\label{field.9}
\end{equation}
the phase function $\vartheta (\alpha ,t)$ is given in (\ref{f.7}). This
result implies an estimate (\ref{mod.13}) of the trace where $\phi (s)$ is
the exponential
\begin{equation}
\phi (s)=\exp \left( -s\right)  \label{field.10}
\end{equation}
and $\zeta (t)$ is the function (\ref{field.9}).

In this first step the identities (\ref{field.8}) and (\ref{field.9}) have
been derived assuming (\ref{field.6}). But under this restriction the
function (\ref{field.9}) is almost periodic. It may grow to large numbers,
but it cannot diverge to infinity. Hence the traces (\ref{field.8}) do not
vanish for $t\rightarrow \infty $. One can achieve a strong decrease which
persists for some finite time interval; but inevitably, recurrences exist.

To derive induced superselection rules one has to violate the condition (\ref
{field.6}). If $h\in \mathcal{D}(M^{-\frac{1}{2}})\setminus \mathcal{D}%
(M^{-1})$ we prove in Appendix \ref{vanHove} that the identities (\ref
{field.8}) and (\ref{field.9}) are still valid. Then an evaluation of (\ref
{field.9}) implies that $\zeta (t)$ diverges for $t\rightarrow \infty $, and
superselection rules follow from (\ref{mod.16}). The time scale of the
decoherence depends only on the vector $h$ in the interaction potential (\ref
{field.2}), and (\ref{field.9}) can increase like $\log t$ or also like $%
t^{\alpha }$ with some $\alpha \in \left( 0,1\right) $, see (\ref{f.13}) and
(\ref{f.14}). The assumption $h\notin \mathcal{D}(M^{-1})$ is therefore
necessary and sufficient for the emergence of superselection rules, which
persist for $t\rightarrow \infty $. Exactly under this condition the Boson
field is known to be infrared divergent. It is still defined on the Fock
space, but the bare Boson number diverges and its ground state disappears in
the continuum, see \cite{Schroer:1963} \cite{Arai/Hirokawa:2000}.

\subsection{Arbitrary normal states as initial state}

The results of Sect. \ref{coherent} can be easily extended to reference
states which are superpositions of a finite number of exponential vectors,
see Appendix \ref{vanHove}. Estimates like (\ref{mod.16}) remain valid with
an additional numerical factor, which increases with the number of
exponential vectors involved. The linear span $\mathcal{L}\left\{ \exp f\mid
f\in \mathcal{H}_{1}\right\} $ of the exponential vectors is a dense linear
subset of the Fock space $\mathcal{H}_{E}=\mathcal{F}(\mathcal{H}_{1})$, and
the convex linear span of all projection operators onto these vectors is a
dense subset $\mathcal{S}_{coh}\subset \mathcal{S}\left( \mathcal{H}%
_{E}\right) $ of all states of the Boson system. We finally obtain for all
reference states $\rho _{E}\in \mathcal{S}_{coh}$ an estimate like (\ref
{mod.16})
\begin{equation}
\left\| P_{S}(\Delta ^{1})\Phi _{t}(\rho _{S})P_{S}(\Delta ^{2})\right\|
_{1}\leq c(\rho _{E})\phi \left( (1-\varepsilon )\delta ^{2}\zeta (t)\right)
,  \label{field.11}
\end{equation}
where $\zeta $ and $\phi $ are again the functions (\ref{field.9}) and (\ref
{field.10}), but with some small $\varepsilon >0$ and an additional
numerical factor $c(\rho _{E})$ which depends on the reference state. If
\newline
$h\in \mathcal{D}(M^{-\frac{1}{2}})\setminus \mathcal{D}(M^{-1})$ this
estimate implies for all $\rho _{E}\in \mathcal{S}_{coh}$%
\begin{equation}
\lim_{t\rightarrow \infty }\left\| P_{S}(\Delta ^{1})\Phi _{t}(\rho
_{S})P_{S}(\Delta ^{2})\right\| _{1}=0  \label{field.12}
\end{equation}
if the intervals $\Delta ^{1}$ and $\Delta ^{2}$ are separated by a distance
$\delta >0$. Due to the factor $c(\rho _{E})$ the emergence of the
superselection sectors $\left\{ P_{S}(\Delta )\mathcal{H}_{S}\right\} $ is
not uniform with respect to $\rho _{E}$; but for suitably restricted subsets
of reference states a fast suppression of the off-diagonal matrix elements
of $\Phi _{t}(\rho _{S})$ can be achieved.

So far we have assumed that the initial state factorizes. The
Schr\"{o}dinger picture allows to start from the more general initial states
\begin{equation}
W=\sum_{\mu =1}^{N}c_{\mu }\,\rho _{S\mu }\otimes \rho _{E\mu }
\label{field.13}
\end{equation}
with $\rho _{S\mu }\in \mathcal{S}(\mathcal{H}_{S}),~\rho _{E\mu }\in
\mathcal{S}_{coh}$ and real (positive and negative) numbers $c_{\mu }$,
which satisfy $\sum_{\mu }c_{\mu }=\mathrm{tr}\,W=1$. Thereby $N$ is an
arbitrary finite number. The set of states (\ref{field.13}) is dense in $%
\mathcal{S}(\mathcal{H}_{S+E})$ and will be denoted by $\mathcal{S}_{fin}(%
\mathcal{H}_{S+E})$. The reduced dynamics for such an initial state
\begin{equation}
\rho _{S}(t)=\widehat{\Phi }_{t}(W):=\mathrm{tr}_{E}U(t)WU^{\dagger }(t)
\label{field.14}
\end{equation}
decomposes into $\rho _{S}(t)=\sum_{\mu }c_{\mu }\Phi _{t}^{\mu }(\rho
_{S\mu })$, where $\Phi _{t}^{\mu }(\,.\,)$ is the reduced dynamics (\ref
{2.2}) with reference state $\rho _{E\mu }$. For all contributions $\Phi
_{t}^{\mu }(\rho _{S\mu })$ estimates of the type (\ref{field.11}) are
valid. Hence $\left\| P_{S}(\Delta ^{1})\widehat{\Phi }_{t}(W)P_{S}(\Delta
^{2})\right\| _{1}\leq \sum_{\mu }\left| c_{\mu }\right| \left\|
P_{S}(\Delta ^{1})\Phi _{t}^{\mu }(\rho _{S\mu })P_{S}(\Delta ^{2})\right\|
_{1}$ implies
\begin{equation}
\lim_{t\rightarrow \infty }\left\| P_{S}(\Delta ^{1})\widehat{\Phi }%
_{t}(W)P_{S}(\Delta ^{2})\right\| _{1}=0  \label{field.15}
\end{equation}
for $W\in \mathcal{S}_{fin}(\mathcal{H}_{S+E})$ and all separated intervals.

By a continuity argument on the mapping (\ref{field.14}) we
finally derive that the superselection sectors $\left\{
P_{S}(\Delta )\mathcal{H}_{S}\mid \Delta \subset
\mathbb{R}\right\} $ emerge for all initial states $W\in
\mathcal{S}(\mathcal{H}_{S+E})$ of the total system.

\begin{theorem}
\label{normal}If the interaction is determined by a vector $h\in \mathcal{D}%
(M^{-\frac{1}{2}})\setminus \mathcal{D}(M^{-1})$ with norm restriction (\ref
{field.4}), then (\ref{field.15}) is true for all initial states $W\in
\mathcal{S}(\mathcal{H}_{S+E})$ and all intervals with distance $\mathrm{dist%
}\left( \Delta ^{1},\Delta ^{2}\right) >0$.
\end{theorem}

\begin{proof}
The mapping (\ref{field.14}) can be extended to a linear mapping $W\in
\mathcal{T}(\mathcal{H}_{S+E})\rightarrow \widehat{\Phi }_{t}(W)\in \mathcal{%
T}(\mathcal{H}_{S})$ which is continuous with respect to the trace norms of
these spaces
\begin{equation}
\left\| \widehat{\Phi }_{t}(W)\right\| _{1}\leq \left\| W\right\| _{1}.
\label{field.16}
\end{equation}
Given some $\varepsilon >0$ and a state $W\in \mathcal{S}(\mathcal{H}_{S+E})$%
, then we can find a $W_{1}\in \mathcal{S}_{fin}(\mathcal{H}_{S+E})$ such
that $\left\| W-W_{1}\right\| _{1}<\varepsilon $. The limit (\ref{field.15})
implies: for intervals $\Delta ^{1}$ and $\Delta ^{2}$ with a non-vanishing
distance there is a time $T(\varepsilon )<\infty $ such that \newline
$\left\| P_{S}(\Delta ^{1})\widehat{\Phi }_{t}(W_{1})P_{S}(\Delta
^{2})\right\| _{1}<\varepsilon $ if $t>T(\varepsilon )$. By linearity of $%
\widehat{\Phi }_{t}$ we have \newline
$\widehat{\Phi }_{t}(W)=\widehat{\Phi }_{t}(W_{1})+\widehat{\Phi }%
_{t}(W-W_{1})$, and we derive the upper bound
\begin{equation}
\begin{array}{l}
\left\| P_{S}(\Delta ^{1})\widehat{\Phi }_{t}(W)P_{S}(\Delta ^{2})\right\|
_{1} \\
\leq \left\| P_{S}(\Delta ^{1})\widehat{\Phi }_{t}(W_{1})P_{S}(\Delta
^{2})\right\| _{1}+\left\| P_{S}(\Delta ^{1})\widehat{\Phi }%
_{t}(W-W_{1})P_{S}(\Delta ^{2})\right\| _{1} \\
\leq \left\| P_{S}(\Delta ^{1})\widehat{\Phi }_{t}(W_{1})P_{S}(\Delta
^{2})\right\| _{1}+\left\| W-W_{1}\right\| _{1}<2\varepsilon
\end{array}
\label{field.17}
\end{equation}
As $\varepsilon $ can be chosen arbitrarily small, the Theorem follows.
\end{proof}

\subsection{KMS states as reference states\label{KMS}}

The considerations presented so far can be extended to an environment with
positive temperature $\beta ^{-1}>0$. That means the Boson field is in a KMS
state\footnote{%
The KMS states of an environment which has a Hamiltonian with a continuous
spectrum cannot be represented by a statistical operator in $\mathcal{S}(%
\mathcal{H}_{E})$. In such a case the algebra of observables has to be
restricted to the Weyl algebra, which is strictly smaller than $\mathcal{B}(%
\mathcal{H}_{E})$, and the KMS states are positive linear functionals on
that algebra.}, which is uniquely characterized by the following expectation
of the Weyl operators $T(h)$%
\begin{equation}
\left\langle T(h)\right\rangle _{\beta }=\exp \left( -\left( h\mid \left( (%
\mathrm{e}^{\beta M}-I)^{-1}+\frac{1}{2}\right) h\right) \right) .
\label{field.21}
\end{equation}

The calculations for a KMS state correspond to the calculations for coherent
states. Only the expectation (\ref{field.20}) has to be substituted by the
expectation (\ref{field.21}). As $(\exp (\beta M)-I)^{-1}$ is a positive
operator we have
\begin{equation}
\left\langle T(h)\right\rangle _{\beta }<\exp \left( -\frac{1}{2}\left\|
h\right\| ^{2}\right) =\left| \mathrm{tr}_{E}T(h)\omega (f)\right| .
\label{field.22}
\end{equation}
Hence in an environment of temperature $\beta ^{-1}>0$ the superselection
sectors are induced on shorter time scale than for coherent states, see
Appendix \ref{vanHove}.

\section{Scattering processes\label{scattering}}

In this final section we investigate the stability of the induced
superselection sectors against additional scattering processes. We restrict
the initial state of the total system to a normal state $W\in \mathcal{S}(%
\mathcal{H}_{S+E})$ to apply standard scattering theory.

The Hamiltonian (\ref{mod.3}) is generalized to
\begin{equation}
H=H_{S\times E}+V,  \label{s.1}
\end{equation}
where $V$ is a scattering potential on $\mathcal{H}_{S\times E}=\mathcal{H}%
_{S}\otimes \mathcal{H}_{E}$. There are no constraints on the commutators $%
\left[ H_{S\times E},V\right] $ or $\left[ F\otimes I_{E},V\right] $, and in
general the dynamics has no conservation law except energy conservation. The
restriction to scattering potentials means that the wave operator $\Omega
=\lim_{t\rightarrow \infty }U^{+}(t)U_{0}(t)$ with $U(t)=\exp (-itH)$ and $%
U_{0}(t)=\exp (-itH_{S\times E})$ exists as strong limit. To simplify the
arguments we assume that there are no bound states and that the wave
operator is unitary on $\mathcal{H}_{S\times E}$. Then the time evolution $%
U(t)=\exp (-itH)$ behaves asymptotically like $U_{0}(t)\Omega ^{+}$ with $%
U_{0}(t)=\exp (-itH_{S\times E})$. More precisely, the existence of wave
operators implies
\begin{equation}
\lim_{t\rightarrow \infty }\,\left\| U(t)WU^{+}(t)-U_{0}(t)\Omega
^{+}W\Omega U_{0}^{+}(t)\right\| _{1}=0  \label{s.2}
\end{equation}
for all $W\in \mathcal{S}(\mathcal{H}_{S+E})$. As $\Omega $ is unitary we
have $\Omega ^{+}W\Omega \in \mathcal{S}(\mathcal{H}_{S\times E})$ for all $%
W\in \mathcal{S}(\mathcal{H}_{S+E})$. Let us denote the reduced dynamics
with the full Hamiltonian (\ref{s.1}) by
\begin{equation}
\rho _{S}(t)=\widehat{\Phi }_{t}(W)=\mathrm{tr}_{E}\left(
U(t)WU^{+}(t)\right) ,  \label{s.3}
\end{equation}
and the reduced dynamics with the Hamiltonian (\ref{mod.3}) by $\widehat{%
\Phi }_{t}^{0}(W)=\mathrm{tr}_{E}\left( U_{0}(t)WU_{0}^{+}(t)\right) $. The
linearity of the trace implies
\begin{equation}
\widehat{\Phi }_{t}(W)=\widehat{\Phi }_{t}^{0}(\Omega ^{+}W\Omega )+\mathrm{%
tr}_{E}\left( U(t)WU^{+}(t)-U_{0}(t)\Omega ^{+}W\Omega U_{0}^{+}(t)\right) .
\label{s.4}
\end{equation}
The off-diagonal contributions of the reduced dynamics (\ref{s.3}) with
scattering can therefore be estimated by
\begin{eqnarray}
\left\| P_{S}(\Delta ^{1})\rho _{S}(t)P_{S}(\Delta ^{2})\right\| _{1} &\leq
&\left\| P_{S}(\Delta ^{1})~\widehat{\Phi }_{t}^{0}(\Omega ^{+}W\Omega
)~P_{S}(\Delta ^{2})\right\| _{1}  \nonumber \\
&&+\left\| U(t)WU^{+}(t)-U_{0}(t)\Omega ^{+}W\Omega U_{0}^{+}(t)\right\|
_{1}.
\end{eqnarray}
As $\Omega ^{+}W\Omega \in \mathcal{S}(\mathcal{H}_{S\times E})$ the first
term vanishes for $t\rightarrow \infty $ under the conditions of Theorem \ref
{normal}, and the second term vanishes for $t\rightarrow \infty $ as a
consequence of (\ref{s.2}). Hence we have derived

\begin{theorem}
If the interaction is determined by a vector $h\in \mathcal{D}(M^{-\frac{1}{2%
}})\setminus \mathcal{D}(M^{-1})$ with norm restriction (\ref{field.4}),
then
\begin{equation}
\lim_{t\rightarrow \infty }\left\| P_{S}(\Delta ^{1})\rho
_{S}(t)P_{S}(\Delta ^{2})\right\| _{1}=0  \label{s.6}
\end{equation}
follows for the reduced dynamics (\ref{s.3}) with the Hamiltonian (\ref{s.1}%
) for all initial states $W\in \mathcal{S}(\mathcal{H}_{S+E})$ and all
intervals with distance $\mathrm{dist}\left( \Delta ^{1},\Delta ^{2}\right)
>0$.
\end{theorem}

Therefore scattering processes do not destroy or modify the
induced superselection sectors $\left\{ P_{S}(\Delta
)\mathcal{H}_{S.}\mid \Delta \subset \mathbb{R}\right\} $, but the
time scale of there emergence increases. Estimates on the time
scale require a more detailed investigation of the scattering
process, which is not given here.

\section{Conclusion}

We have investigated a class of systems, which are coupled to a mass zero
Boson field. These models exhibit the following properties:

\begin{itemize}
\item  The Boson field induces superselection rules into the system, if and
only if the field is infrared divergent. Thereby infrared divergence means
that the bare Boson number diverges and the Boson vacuum disappears in the
continuum, but the Hamiltonian remains bounded from below.

\item  The superselection sectors are fully determined by the Hamiltonian,
they finally emerge for all normal initial states of the total system, --
including non-product states -- and for KMS states as reference states of
the Boson system.

\item  The time scale of the decoherence depends on the interaction and on
the initial state. There are restrictions on the reference state of the
Boson field to obtain superselection rules, which are effective within a
short time.

\item  The superselection sectors persist, if additional scattering
processes take place. In this case the total system may have no conservation
law except energy conservation.
\end{itemize}

These results underline the known importance of low frequency excitations of
the environment for the process of decoherence \cite{LCD:1987} \cite
{DellAntonio:2003}. \appendix

\section{Estimates of operators\label{estimates}}

Let $P:\Delta =\left[ a,b\right) \subset \mathbb{R}\rightarrow \mathcal{L}(%
\mathcal{H})$ be a family of orthogonal projectors in $\mathcal{H}$ with the
properties (\ref{2.4}). The mapping $P(\Delta )$ can be extended to a $%
\sigma $-additive measure on the Borel algebra
$\mathcal{B}(\mathbb{R})$
generated by open subsets of the real line $\mathbb{R}$. The operators $P\left( %
\left[ a,b\right) \right) $ are naturally left continuous in both variables $%
a$ and $b$. In what follows we investigate some integrals of bounded
operator-valued functions with respect to $P$. More details can be found in
\cite{Kupsch/Smolyanov:2004}.

\begin{lemma}
\label{p1} Let $f:\mathbb{R}\rightarrow \mathcal{T}(\mathcal{H})$
be a differentiable function with a Bochner integrable derivative
$f^{\prime }(x)\in \mathcal{T}(\mathcal{H})$. Then for any
interval $\Delta =\left[ a,b\right) \subset \mathbb{R}$ the
following identity holds
\begin{equation}
\begin{array}{r}
\int_{a}^{b}P(dx)f(x)=P(\left[ a,b\right) )f(b)-\int_{a}^{b}P(\left[
a,x\right) )f^{\prime }(x)dx \\
=P(\left[ a,b\right) )f(a)+\int_{a}^{b}P(\left[ x,b\right) )f^{\prime }(x)dx,
\end{array}
\label{a1}
\end{equation}
and the norm of this integral has the upper bound
\begin{equation}
\left\| \int_{\Delta }P(dx)f(x)\right\| \leq \min \left( \left\|
f(a)\right\| ,\left\| f(b)\right\| \right) +\int_{\Delta }\left\| f^{\prime
}(x)\right\| _{1}dx.  \label{a2}
\end{equation}
\end{lemma}

\begin{proof}
The identities (\ref{a1}) are just the integration by parts formula of the
Stieltjes integral, see e.g. \cite{Baumg/Wollenberg:1983} Sect. 5.1. The
norm estimate (\ref{a2}) is then a consequence of $\left\| P(\Delta
)\right\| \leq 1$ and the rule $\left\| AB\right\| _{1}\leq \left\|
A\right\| \left\| B\right\| _{1}$ for the trace norm.
\end{proof}

The same type of identities and norm estimates can be derived for integrals
\newline
$\int_{\Delta }f(x)P(dx)=\left( \int_{\Delta }P(dx)f^{+}(x)\right) ^{+}$
with a reversed order of the operators. An immediate consequence of Lemma
\ref{p1} is the

\begin{corollary}
\label{c1} Let $f:\mathbb{R}\rightarrow \mathcal{T}(\mathcal{H})$
be a function with a Bochner integrable derivative \newline
$f^{\prime }(x)\in \mathcal{T}(\mathcal{H})$. If $\left\|
f(x)\right\| _{1}$ vanishes for $x\rightarrow \pm \infty $ the
identities
\begin{equation}
\begin{array}{l}
\int_{a}^{\infty }P(dx)f(x)=-\int_{a}^{\infty }P(\left[ a,x\right)
)f^{\prime }(x)dx\quad \mathrm{and} \\
\int_{-\infty }^{b}P(dx)f(x)=\int_{-\infty }^{b}P(\left[ x,b\right)
)f^{\prime }(x)dx
\end{array}
\label{a3}
\end{equation}
hold for all $a,b\in \mathbb{R}$ and the estimate
\begin{equation}
\left\| \int_{\Delta }P(dx)f(x)\right\| _{1}\leq \int_{\Delta }\left\|
f^{\prime }(x)\right\| _{1}dx.  \label{a4}
\end{equation}
follows for the infinite intervals $\Delta =\left[ a,\infty \right) $ and $%
\left( -\infty ,b\right) $.
\end{corollary}

We now consider operators
\begin{equation}
S_{\varphi }=\int_{\mathbb{R}\times \mathbb{R}}\varphi
(x,y)P(dx)SP(dy)  \label{a5}
\end{equation}
where $S\in \mathcal{T}(\mathcal{H})$ and $\varphi :\mathbb{R}\times \mathbb{R}%
\rightarrow \mathbb{C}$ is a differentiable function. We obviously have $%
P(\Delta ^{1})S_{\varphi }P(\Delta ^{2})=\int_{\Delta ^{1}\times \Delta
^{2}}\varphi (x,y)P(dx)SP(dy)$. First let us notice that
\begin{equation}
P(\Delta ^{1})S_{\varphi }P(\Delta ^{2})=\int_{\Delta
^{1}}P(dx)S\int_{\Delta ^{2}}\varphi (x,y)P(dy)=\int_{\Delta ^{1}}P(dx)SA(x),
\label{a6}
\end{equation}
where the function $A(x)$ is defined by $A(x)=\int_{\Delta ^{2}}\varphi
(x,y)P(dy)\in \mathcal{L}(H)$. Its derivative is $A^{\prime
}(x)=\int_{\Delta ^{2}}\varphi _{1}(x,y)P(dy)$ with $\varphi _{1}(x,y)=\frac{%
\partial }{\partial x}\varphi (x,y)$. The operator norm of this derivative
has the upper bound $\left\| A^{\prime }(x)\right\| \leq \sup_{y\in \Delta
^{2}}\left| \varphi _{1}(x,y)\right| $. Then (\ref{a6}) can be estimated by
Corollary \ref{c1}. We formulate the final result for operators (\ref{a4}) $%
S_{\varphi }$ with a function $\varphi (x,y)=\chi (x-y)$ which depends only
on the difference $x-y$.

\begin{theorem}
Let $\chi :x\in \mathbb{R}\rightarrow \mathbb{C}$ be a
differentiable complex-valued function with $\chi (x)\rightarrow
0$ if $\left| x\right| \rightarrow \infty $ and $\left| \chi
^{\prime }(x)\right| \leq \phi (\left| x\right| )$, where $\phi
(s)$ is non-increasing for $s\geq 0$ with a bounded integral
$\int_{0}^{\infty }\phi (x)dx<\infty $. Then for any nuclear
operator $S$ the operator $S_{\varphi }$ with $\varphi (x,y)=\chi
(x-y)$ is again nuclear, and for the disjoint intervals $\Delta
^{1}=\left( -\infty ,b_{1}\right) $ and $\Delta ^{2}=\left[
a_{2},\infty \right) $ with $\delta =a_{2}-b_{1}\geq 0$ the
following estimate holds
\begin{equation}
\left\| P(\Delta ^{1})S_{\varphi }P(\Delta ^{2})\right\| _{1}\leq \left\|
S\right\| _{1}\int_{\delta }^{\infty }\phi (x)dx.  \label{a19}
\end{equation}
\end{theorem}

\section{The van Hove model\label{vanHove}}

\subsection{The Hamiltonian}

Let $F\circ G$ denote the symmetric tensor product of the Fock space $%
\mathcal{F}(\mathcal{H}_{1})$ with vacuum $1_{vac}$. For all $f\in \mathcal{H%
}_{1}$ the exponential vectors $\exp f=1_{vac}+f+\frac{1}{2}f\circ f+...$
converge within $\mathcal{F}(\mathcal{H}_{1})$, the inner product being $%
\left( \exp f\mid \exp g\right) =\exp \left( f\mid g\right) $. Coherent
vectors (states) are the normalized exponential vectors $\exp \left( f-\frac{%
1}{2}\left\| f\right\| ^{2}\right) $. The linear span of all exponential
vectors $\left\{ \exp f\mid f\in \mathcal{H}_{1}\right\} $ is dense in $%
\mathcal{F}(\mathcal{H}_{1})$. The creation operators $a^{+}(f)$ are
uniquely determined by $a^{+}(f)\exp g=f\circ \exp g=\newline
\frac{\partial }{\partial \lambda }\exp (f+\lambda g)\mid _{\lambda =0}$
with $f,g\in \mathcal{H}_{1}$ and the annihilation operators are given by $%
a(g)\exp f=\left( g\mid f\right) \exp f$. These operators satisfy the
standard commutation relations $\left[ a(f),a^{+}(g)\right] =\left( f\mid
g\right) $. If $M$ is a operator on $\mathcal{H}_{1}$ then $\Gamma (M)$ is
uniquely defined as operator on $\mathcal{F}(\mathcal{H}_{1})$ by $\Gamma
(M)\exp f:=\exp (Mf)$, and the derivation $d\Gamma (M)$ is defined by $%
d\Gamma (M)\exp f:=(Mf)\circ \exp f$.

For arbitrary elements $g\in \mathcal{H}_{1}$ the unitary Weyl operators are
defined on the set of exponential vectors by $T(g)\exp f=\exp \left( -\left(
g\mid f\right) -\frac{1}{2}\left\| g\right\| ^{2}\right) \exp (f+g)$. This
definition is equivalent to $T(g)=\exp \left( a^{+}(g)-a(g)\right) $. The
Weyl operators are characterized by the properties
\begin{equation}
\begin{array}{c}
T(g_{1})T(g_{2})=T(g_{1}+g_{2})\,\exp \left( -i\,\mathrm{Im}\left( g_{1}\mid
g_{2}\right) \right) \\
\left( 1_{vac}\mid T(g)\,1_{vac}\right) =\exp \left( -\frac{1}{2}\left\|
g\right\| ^{2}\right) .
\end{array}
\label{f.0}
\end{equation}
The matrix element of $T(h)$ between coherent vectors $\exp \left( f-\frac{1%
}{2}\left\| f\right\| ^{2}\right) =T(f)1_{vac}$ follows from these relations
as
\begin{equation}
\left( 1_{vac}\mid T^{+}(g)T(h)T(f)\,1_{vac}\right) =\exp \left( -\frac{1}{2}%
\left\| h+f-g\right\| ^{2}+i\,\mathrm{Im}\left\{ \left( g\mid f\right)
+\left( f+g\mid h\right) \right\} \right) .  \label{f.1}
\end{equation}
For a free field the time evolution on the Fock space is given by $U(t)=\exp
(-iH_{E}t)=\Gamma \left( V(t)\right) $ with $V(t):=\exp (-iMt)$. For
exponential vectors we obtain $U(t)\exp f=\exp \left( V(t)f\right) $. From
these equations the dynamics of the Weyl operators follows as
\begin{equation}
U^{+}(t)T(g)U(t)=T\left( V^{+}(t)\,g\right) .  \label{f.2}
\end{equation}
For fixed $h\in \mathcal{H}_{1}$ the unitary operators $T^{+}(h)U(t)T(h),\,t%
\in \mathbb{R}$, form a one parameter group which acts on exponential vectors as%
\newline
$T^{+}(h)U(t)T(h)\exp f=\exp \left( \left( h\mid V(t)(f+h)-f\right) -\left\|
h\right\| ^{2}\right) \exp \left( V(t)(f+h)-h\right) $. \newline
For $h\in \mathcal{H}_{1}$ with $Mh\in \mathcal{H}_{1}$ the generator of
this group is easily identified with \newline
$T^{+}(h)H_{E}T(h)=\,H_{E}+\Phi (Mh)+\left( h\mid Mh\right) $, where $\Phi
(.)$ is the field operator. This identity was first derived by Cook \cite
{Cook:1961} by quite different methods. If $h$ satisfies $M^{-1}h\in
\mathcal{H}_{1}$ we obtain
\begin{equation}
T^{+}(M^{-1}h)H_{E}T(M^{-1}h)-\left\| M^{-\frac{1}{2}}h\right\|
^{2}=H_{E}+\Phi (h)  \label{f.3}
\end{equation}
which is the Hamiltonian of the van Hove model \cite{Hove:1952}, see also,
\cite{Berezin:1966} p.166ff, \cite{Emch:1972} and \cite{Arai/Hirokawa:2000}.

For all $h\in \mathcal{H}_{E}$ with $M^{-\frac{1}{2}}h\in \mathcal{H}_{E}$
the field operator $\Phi (h)$ satisfies the estimate
\begin{equation}
\left\| \Phi (h)\psi \right\| \leq 2\left\| M^{-\frac{1}{2}}h\right\|
\left\| \sqrt{H_{E}}\psi \right\| +\left\| h\right\| \left\| \psi \right\| ,
\label{f.4}
\end{equation}
where $\psi \in \mathcal{F}(\mathcal{H}_{1})$ is an arbitrary vector in the
domain of $H_{E}$, see e.g. eq. (2.3) of \cite{Arai/Hirokawa:1997}. As
consequences we obtain the following Lemma for the Hamiltonian of the van
Hove model, see \cite{Schroer:1963} and \cite{Arai/Hirokawa:2000}.

\begin{lemma}
\label{Selfadjoint}The operators $H_{E}+\lambda \Phi (h),\,\lambda \in \mathbb{R%
}$, are self-adjoint on the domain of $H_{E}$ if $h\in \mathcal{D}(M^{-\frac{%
1}{2}})$.
\end{lemma}

\begin{proof}
From (\ref{f.4}) and the numerical inequality $\sqrt{x}\leq ax+(4a)^{-1}$,
valid for $x\geq 0$ and $a>0$, we obtain a bound $\left\| \Phi (h)\psi
\right\| \leq c_{1}\left\| H_{E}\psi \right\| +c_{2}\left\| \psi \right\| $
with positive numbers $c_{1},\,c_{2}>0$ where $c_{1}$ can be chosen
arbitrarily small. Then the Kato-Rellich Theorem yields the first statement.
\end{proof}

A further consequence is

\begin{lemma}
\label{bound} The operator $H_{E}-\frac{1}{2}\Phi ^{2}(h)$ has the lower
bound $H_{E}-\frac{1}{2}\Phi ^{2}(h)\geq -\left\| h\right\| ^{2}$, if $h\in
\mathcal{H}_{1}$ and $\left\| M^{-\frac{1}{2}}h\right\| \leq 2^{-1}$.
\end{lemma}

\begin{proof}
From (\ref{f.4}) we obtain\newline
$\left\| \Phi (h)\psi \right\| ^{2}\leq 4\left\| M^{-\frac{1}{2}}h\right\|
^{2}\left( \psi \mid H_{E}\psi \right) +4\left\| M^{-\frac{1}{2}}h\right\|
\left\| h\right\| \left\| \sqrt{H_{E}}\psi \right\| \left\| \psi \right\|
+\left\| h\right\| ^{2}\left\| \psi \right\| ^{2}\newline
\leq 8\left\| M^{-\frac{1}{2}}h\right\| ^{2}\left( \psi \mid H_{E}\psi
\right) +2\left\| h\right\| ^{2}\left\| \psi \right\| ^{2}.$ Hence the
operator inequalities \newline
$0\leq \frac{1}{2}\Phi ^{2}(h)\leq 4\left\| M^{-\frac{1}{2}}h\right\|
^{2}H_{E}+\left\| h\right\| ^{2}I_{E}$ hold, and Lemma \ref{bound} follows.
\end{proof}

Therefore the total Hamiltonian (\ref{mod.3}) is semibounded, and the
unitary operators \newline
$U_{\lambda }(t)=\exp \left( -i(H_{E}+\lambda \Phi (h))t\right) $ are well
defined if (\ref{field.4}) is satisfied.

\subsection{Evaluation of the traces}

In a first step we evaluate the expectation value of (\ref{field.5}) $%
U_{\alpha \beta }(t)=U_{\alpha }(-t)U_{\beta }(t)$ for a coherent state (=
normalized exponential vector) $\exp \left( f-\frac{1}{2}\left\| f\right\|
^{2}\right) =T(f)\,1_{vac}$ under the additional constraint $h\in \mathcal{D}%
(M^{-1})$. This assumption allows to use the identity (\ref{f.3}) which
reduces all calculations to the Weyl relations and the vacuum expectation (%
\ref{f.1}). The extension to the general case, which violates $h\in \mathcal{%
D}(M^{-1})$, can then be performed by a continuity argument.

If $M^{-1}h\in \mathcal{H}_{1}$ the identity (\ref{f.3}) implies \newline
$U_{\lambda }(t)=T(-\lambda M^{-1}h)U_{0}(t)T(\lambda M^{-1}h)\exp \left(
i\lambda ^{2}\left( h\mid M^{-1}h\right) t\right) $. Then $U_{\alpha \beta
}(t)=U_{\alpha }(-t)U_{\beta }(t)$ can be calculated with the help of (\ref
{f.0}) and (\ref{f.2}) as
\begin{equation}
\begin{array}{l}
U_{\alpha \beta }(t)=T\left( (\alpha -\beta )\left( V^{+}(t)-I\right)
M^{-1}h\right) \,\exp \left( -i\eta (t)\right) , \\
\eta (t)=-(\alpha ^{2}-\beta ^{2})\left\{ \left( h\mid M^{-1}h\right)
t-\left( M^{-1}h\mid M^{-1}\sin (Mt)h\right) \right\} .
\end{array}
\label{f.5}
\end{equation}
The matrix element of $U_{\alpha \beta }(t)$ between the coherent states $%
T(f)\,1_{vac}$ and $T(g)\,1_{vac}$ is then evaluated with the help of (\ref
{f.1})
\begin{equation}
\begin{array}{r}
\left( 1_{vac}\mid T^{+}(g)U_{\alpha \beta }(t)T(f)\,1_{vac}\right) =\exp
\left( -\frac{1}{2}\left\| \left( \alpha -\beta \right) \left(
V^{+}(t)-I\right) M^{-1}h+f-g\right\| ^{2}\right) \\
\times \exp \left( i\,\mathrm{Im}(g\mid f)\right) \times \exp \left( i\left(
\vartheta (\alpha ,t)-\vartheta (\beta ,t)\right) \right)
\end{array}
\label{f.6}
\end{equation}
with the phase function
\begin{equation}
\vartheta (\alpha ,t)=-\alpha \,\mathrm{Im}\left( f+g\mid \left(
I-V^{+}(t)\right) M^{-1}h\right) -\alpha ^{2}\left( M^{-1}h\mid
ht-M^{-1}\sin (Mt)h\right) .  \label{f.7}
\end{equation}
For $g=f$ the identity (\ref{f.6}) leads to the trace (\ref{field.8}).

So far we have assumed $h\in \mathcal{D}(M^{-1})$. Then the norm $\left\|
\left( V^{+}(t)-I\right) M^{-1}h\right\| =\newline
\left\| \left( I-\exp \left( iMt\right) \right) M^{-1}h\right\| $ is an
almost periodic function of $t$, and induced superselection rule can emerge
only in an approximate sense on an intermediate time scale. But $\left(
V^{+}(t)-I\right) M^{-1}h$ is a vector in $\mathcal{H}_{1}$ also under the
weaker condition $h\in \mathcal{D}(M^{-\frac{1}{2}})\supset \mathcal{D}%
(M^{-1})$. Moreover from Lemma \ref{Selfadjoint} we know that the van Hove
Hamiltonians $H_{E}+\lambda \Phi (h)$ and the groups $U_{\lambda }(t)$ are
defined under this weaker condition. In the next step we shall use a
continuity argument to prove that (\ref{f.6}) is indeed still valid for
vectors $h\in \mathcal{D}(M^{-\frac{1}{2}})$ without knowing whether $h\in
\mathcal{D}(M^{-1})$ or not. Then we derive the essential statement that the
norm $\left\| \left( V^{+}(t)-I\right) M^{-1}h\right\| $ diverges for $%
t\rightarrow \infty $ if $h\notin \mathcal{D}(M^{-1}).$ As this behaviour is
possible under the condition (\ref{field.4}), which guarantees the existence
of a semibounded Hamiltonian (\ref{mod.3}), stable superselection sectors
emerge if $h$ is chosen such that $h\in \mathcal{D}(M^{-\frac{1}{2}%
})\setminus \mathcal{D}(M^{-1})$ with the additional constraint (\ref
{field.4}).

For the proof of this statement we introduce the norm
\begin{equation}
\left| \left\| h\right\| \right| :=\left\| h\right\| +\left\| M^{-\frac{1}{2}%
}h\right\| .  \label{f.9}
\end{equation}
Let $h_{n}\in \mathcal{H}_{1},\,n=1,2,...,$ be a sequence of real vectors
which converges in this topology to a vector $h$, then we know from (\ref
{f.4}) and the proof of Lemma \ref{Selfadjoint} that there exist two null
sequences of positive numbers $c_{1n}$ and $c_{2n}$ such that
\[
\left\| \left( \Phi (h_{n})-\Phi (h)\right) \psi \right\| \leq c_{1n}\left\|
\left( H_{E}+\Phi (h)\right) \psi \right\| +c_{2n}\left\| \psi \right\| .
\]
Hence the operators $H_{E}+\Phi (h_{n})$ converge strongly to $H_{E}+\Phi
(h) $ and the groups $U(h_{n};t)=\exp \left( -i\left( H_{E}+\Phi
(h_{n})\right) t\right) $ converge strongly to the group $U(h;t)=\exp \left(
-i\left( H_{E}+\Phi (h)\right) t\right) $, uniformly in any finite interval $%
0\leq t\leq s<\infty $; see e.g. Theorem 4.4 on p. 82 of \cite{Maslov:1972},
or Theorem 3.17 of \cite{Davies:1980}. The operators \newline
$U_{\alpha \beta ,n}(t):=\exp \left( i\left( H_{E}+\alpha \Phi
(h_{n})\right) t\right) \exp \left( -i\left( H_{E}+\beta \Phi (h_{n})\right)
t\right) $ converge therefore in the weak operator topology to $U_{\alpha
\beta }(t)$. For $n=1,2,..$ we can calculate the corresponding traces $%
\mathrm{tr}_{E}U_{\alpha \beta ,n}(t)\omega (f)$ with the result (\ref
{field.9}), where $h$ has to be substituted by $h_{n}$. Since (\ref{field.9}%
) is continuous in the variable $h$ in the topology (\ref{f.9}) the limit
for $n\rightarrow \infty $ is again given by (\ref{field.9}).

To prove the divergence of $\left\| \left( V^{+}(t)-I\right) M^{-1}h\right\|
$ for $t\rightarrow \infty $ we introduce the spectral resolution $%
P_{M}(d\lambda )$ of the one-particle Hamilton operator $M$. The energy
distribution of the vector $h\in \mathcal{H}_{1}$ is given by the measure $%
d\sigma _{h}(\lambda )=\left( h\mid P_{M}(d\lambda )h\right) $. The exponent
(\ref{field.9}) is then the integral
\begin{equation}
\zeta (t)=\frac{1}{2}\left\| \left( I-\exp \left( iMt\right) \right)
M^{-1}h\right\| ^{2}=2\int_{\mathbb{R}_{+}}\lambda ^{-2}\sin ^{2}\frac{\lambda t%
}{2}\,d\sigma _{h}(\lambda ).  \label{f.10}
\end{equation}
This integral is well defined for all $h\in \mathcal{H}_{1}$, and
$\zeta (t)$ is a differentiable function for $t\in \mathbb{R}$.
The requirement $h\in \mathcal{D}(M^{-\frac{1}{2}})\setminus
\mathcal{D}(M^{-1})$ is equivalent to the conditions
$\int_{0}^{\infty }\lambda ^{-1}d\sigma _{h}(\lambda )<\infty $
and
\begin{equation}
\int_{\varepsilon }^{\infty }\lambda ^{-2}\,d\sigma _{h}(\lambda )\nearrow
\infty \quad \mathrm{if}\;\varepsilon \rightarrow +0.  \label{f.11}
\end{equation}

\begin{lemma}
If $h\notin \mathcal{D}(M^{-1})$, i.e. (\ref{f.11}), the integral (\ref{f.10}%
) diverges for $t\rightarrow \infty $.
\end{lemma}

\begin{proof}
Since the operator $M$ has an absolutely continuous spectrum, the measure $%
d\sigma _{h}(\lambda )$ is absolutely continuous with respect to the
Lebesgue measure $d\lambda $ on $\mathbb{R}_{+}$. Consequently, the measure $%
\lambda ^{-2}\,d\sigma _{h}(\lambda )$ is absolutely continuous with respect
to the Lebesgue measure on any interval $\left( \varepsilon ,\infty \right) $
with $\varepsilon >0$. The identity $\sin ^{2}\frac{\lambda t}{2}=\frac{1}{2}%
\left( 1-\cos \lambda t\right) $ and the Lebesgue Lemma therefore imply
\newline
$\lim_{t\rightarrow \infty }\int_{\varepsilon }^{\infty }\lambda ^{-2}\sin
^{2}\frac{\lambda t}{2}\,d\sigma _{h}(\lambda )=\frac{1}{2}\int_{\varepsilon
}^{\infty }\lambda ^{-2}\,d\sigma _{h}(\lambda )$. Given a number $\Lambda
>0 $ the assumption (\ref{f.11}) yields the existence of an $\varepsilon >0$
such that
\begin{equation}
\lim_{t\rightarrow \infty }\int_{\varepsilon }^{\infty }\lambda ^{-2}\sin
^{2}\frac{\lambda t}{2}\,d\sigma _{h}(\lambda )=\frac{1}{2}\int_{\varepsilon
}^{\infty }\lambda ^{-2}\,d\sigma _{h}(\lambda )>\Lambda .  \label{f.12}
\end{equation}
From the inequality $\int_{\mathbb{R}_{+}}\lambda ^{-2}\sin ^{2}\frac{\lambda t%
}{2}\,d\sigma _{h}(\lambda )\geq \int_{\varepsilon }^{\infty }\lambda
^{-2}\sin ^{2}\frac{\lambda t}{2}\,d\sigma _{h}(\lambda )$ we then obtain%
\newline
$\int_{0}^{\infty }\lambda ^{-2}\sin ^{2}\frac{\lambda t}{2}\,d\sigma
_{h}(\lambda )>\Lambda $ for sufficiently large $t$. Since the number $%
\Lambda $ can be arbitrarily large the integral (\ref{f.10}) diverges for $%
t\rightarrow \infty $.
\end{proof}

If $d\sigma _{h}(\lambda )$ satisfies additional regularity conditions, we
can obtain more precise statements. A powerlike behaviour $d\sigma
_{h}(\lambda )\cong c\cdot \lambda ^{2\mu }d\lambda ,\,c>0$, near $\lambda
=+0$ is compatible with the requirement $h\in \mathcal{D}(M^{-\frac{1}{2}%
})\setminus \mathcal{D}(M^{-1})$ if $0<\mu \leq \frac{1}{2}$. For the ohmic
case $d\sigma _{h}(\lambda )\cong c\cdot \lambda d\lambda $ we obtain
\begin{eqnarray}
\zeta (t) &=&2\int_{0}^{\infty }\lambda ^{-2}\sin ^{2}\frac{\lambda t}{2}%
\,d\sigma _{h}(\lambda )=\int_{0}^{\infty }\lambda ^{-2}\left( 1-\cos
\lambda t\right) \,d\sigma _{h}(\lambda )  \nonumber \\
&\simeq &c\int_{0}^{t}s^{-1}\left( 1-\cos s\right) \,ds\simeq c\log t\quad
\mathrm{for}\;t\rightarrow \infty ;  \label{f.13}
\end{eqnarray}
and the subohmic case $d\sigma _{h}(\lambda )\cong c\cdot \lambda ^{2\mu
}d\lambda $ with $0<\mu <\frac{1}{2}$ implies a powerlike divergence
\begin{eqnarray}
\zeta (t) &=&\int_{0}^{\infty }\lambda ^{-2}\left( 1-\cos \lambda t\right)
\,d\sigma _{h}(\lambda )  \nonumber \\
&\simeq &c\,t^{1-2\mu }\int_{0}^{t}s^{-2+2\mu }\left( 1-\cos s\right)
\,ds\sim t^{1-2\mu }\quad \mathrm{for}\;t\rightarrow \infty .  \label{f.14}
\end{eqnarray}

So far the reference state $\omega $ has been a coherent state. But the
results remain true if we take as reference state $\omega $ the projection
onto a vector $\psi =\sum_{n=1}^{N}c_{n}\exp f_{n},\,f_{n}\in \mathcal{H}%
_{1} $, which is a finite linear combination of exponential vectors. In that
case the trace (\ref{mod.11}) is a sum of terms (\ref{f.6}) with $f$ and $g$
given by the vectors $f_{n}$. The exponent $\zeta _{\alpha \beta }\left[ f,g%
\right] (t):=\frac{1}{2}\left\| \left( \alpha -\beta \right) \left(
V^{+}(t)-I\right) M^{-1}h+f-g\right\| ^{2}$ in (\ref{f.6}) diverges for $%
\alpha \neq \beta $ under the same condition as (\ref{f.10}) does. Moreover,
the asymptotic behaviour of $\zeta _{\alpha \beta }\left[ f,g\right] (t)$ is
dominated by the asymptotics of $\left( \alpha -\beta \right) ^{2}\zeta (t)$%
. Hence uniform estimates like (\ref{mod.16}) remain valid, but the function
$\phi \left( \delta ^{2}\zeta (t)\right) $ should be substituted by $c\cdot
\phi \left( (1-\varepsilon )\delta ^{2}\zeta (t)\right) $ with a small $%
\varepsilon >0$, and a constant $c\geq 1$, which depends on the coefficients
$c_{n}$ and on the norms $\left\| f_{m}-f_{n}\right\| $. This factor
increases with the number $N$ of the exponential vectors.

In the case of a KMS state of temperature $\beta ^{-1}>0$ the calculations
essentially follow the calculations for coherent states. The expectation of $%
U_{\mu \nu }(t)$ is calculated using (\ref{field.21}). The result
\begin{equation}
\left\langle U_{\mu \nu }(t)\right\rangle _{\beta }=\exp \left( -\left( \mu
-\nu \right) ^{2}\zeta _{\beta }(t)\right) \exp \left( i\left( \vartheta
(\mu ,t)-\vartheta (\nu ,t)\right) \right)  \label{f.15}
\end{equation}
has the same structure as (\ref{field.8}) with the temperature dependent
function
\begin{eqnarray}
\zeta _{\beta }(t) &=&\left( \left( I-\mathrm{e}^{Mt}\right) M^{-1}h\mid
\left( (\mathrm{e}^{\beta M}-I)^{-1}+\frac{1}{2}\right) \left( I-\mathrm{e}%
^{Mt}\right) M^{-1}h\right)  \nonumber \\
&\geq &\frac{1}{2}\left\| \left( I-\exp \left( Mt\right) \right)
M^{-1}h\right\| ^{2}=\zeta (t),  \label{f.16}
\end{eqnarray}
and the phase function $\vartheta (\mu ,t)=-\mu ^{2}\left( M^{-1}h\mid
ht+M^{-1}\sin (Mt)h\right) $, which originates from (\ref{f.5}). The
inequality (\ref{f.16}) implies that for $h\in \mathcal{D}(M^{-\frac{1}{2}%
})\setminus \mathcal{D}(M^{-1})$ superselection sectors are induced on a
shorter time scale than for coherent states.

As a final remark we indicate a modification of the model, which does not
use the absolute continuity of the spectrum of $M$. But we still need a
dominating low energy contribution in the interaction. More precisely, we
assume that $\sigma _{h}(\lambda )\equiv \int_{0}^{\lambda }d\sigma
_{h}(\alpha )$ behaves at low energies like
\begin{equation}
\lambda ^{-2}\sigma _{h}(\lambda )\nearrow \infty \;\mathrm{if}\;\lambda
\rightarrow +0.  \label{f.17}
\end{equation}
Then we can derive the divergence of (\ref{f.10}) by the inequalities\newline
$\zeta (t)\geq 4\int_{\mathbf{0}}^{\frac{\pi }{t}}\lambda ^{-2}\sin ^{2}%
\frac{\lambda t}{2}\,d\sigma _{h}(\lambda )\geq \frac{4}{\pi ^{2}}t^{2}\int_{%
\mathbf{0}}^{\frac{\pi }{t}}\,d\sigma _{h}(\lambda )=\frac{4}{\pi ^{2}}%
t^{2}\,\sigma _{h}(\frac{\pi }{t})$ using $\sin x\geq \frac{2}{\pi }x$ if $%
0\leq x\leq \frac{\pi }{2}$. For measures $d\sigma _{h}(\lambda )\sim
\lambda ^{2\mu }d\lambda $ the assumption (\ref{f.17}) is more restrictive
than (\ref{f.11}) -- it excludes $d\sigma _{h}(\lambda )\sim \lambda
d\lambda $ which satisfies the conditions of Lemma 2. But (\ref{f.17}) is
also meaningful for point measures $d\sigma _{h}(\lambda )$, and $M$ may be
an operator with a pure point spectrum. The Boson field can therefore be
substituted by an infinite family of harmonic oscillators, which have zero
as accumulation point of their frequencies. Such an example has been
discussed -- also for KMS states -- by Primas \cite{Primas:2000}.

\end{document}